\documentclass[preprint,preprintnumbers,nobibnotes,prb,amsmath,amssymb]{revtex4}   
\usepackage{graphicx} 
\usepackage{epsfig}

\newcommand  {\etal}     {{\it et al.}}

\newcommand  {\Biopol}   {{\it Biopolymers\ }}

\newcommand  {\BJ}       {{\it Biophys.~J.\ }}

\newcommand  {\COSB}     {{\it Curr.\ Opin.\ Struct.\ Biol.\ }}

\newcommand  {\JBP}      {{\it J.\ Biol.\ Phys.\ \ }}
\newcommand  {\JCB}      {{\it J.\ Comp.\ Biol.\ \ }}

\newcommand  {\JCP}      {{\it J.\ Chem.\ Phys.\ }}
\newcommand  {\JMB}      {{\it J.\ Mol.\ Biol.\ }}
\newcommand  {\JMGM}     {{\it J.\ Mol.\ Graphics\ Modell.\ }}

\newcommand  {\Mac}      {{\it Macromolecules\ }}

\newcommand  {\Pro}      {{\it Proteins\ }}

\newcommand  {\ProSci}   {{\it Protein\ Sci.\ }}

\newcommand  {\PNAS}     {{\it Proc.\ Natl.\ Acad.\ Sci.\ USA\ }}

\newcommand  {\PRE}      {{\it Phys.\ Rev.\ E\ }}
\newcommand  {\PRL}      {{\it Phys.\ Rev.\ Lett.\ }}

\newcommand  {\SFD}      {{\it Struct.\ Fold.\ Des.\ }}

\newcommand{\beq}{\begin{equation}}
\newcommand{\eeq}{\end{equation}}
\newcommand{\beqa}{\begin{eqnarray}}
\newcommand{\eeqa}{\end{eqnarray}}
\newcommand{\bea}{\begin{eqnarray}}
\newcommand{\eea}{\end{eqnarray}}
\newcommand   {\ev}[1]   {\langle #1\rangle}

\newcommand   {\Pb}      {P_{\mbox{{\scriptsize b}}}}
\newcommand   {\Pd}      {P_{\mbox{{\scriptsize d}}}}
\newcommand   {\Pf}      {P_{\mbox{{\scriptsize f}}}}
\newcommand   {\Qp}      {Q^{\prime}}
\newcommand   {\eb}      {\epsilon_{\mbox{{\scriptsize b}}}}
\newcommand   {\Zspa}    {Z${}_{{\rm SPA-1}}$}

\begin{document}

\preprint{LU TP 03-44}

\title{Coupled Folding-Binding versus Docking:\\ 
A Lattice Model Study}

\author{Nitin Gupta}
\email{nitingpt@iitk.ac.in}   
\affiliation{
Department of Computer Science and Engineering\\
Indian Institute of Technology Kanpur, Kanpur-208016, India}

\author{Anders Irb\"ack}
\email{anders@thep.lu.se}   
\affiliation{Complex Systems Division, Department of Theoretical Physics\\
Lund University,  S\"olvegatan 14A,  SE-223 62 Lund, Sweden}

\date{November 24, 2003}

\begin{abstract}
Using a simple hydrophobic/polar protein model, we perform a 
Monte Carlo study of the thermodynamics and kinetics of binding 
to a target structure for two closely related sequences, 
one of which has a unique folded state while the other 
is unstructured. We obtain significant differences in their binding behavior.
The stable sequence has rigid docking as its preferred binding mode, 
while the unstructured chain tends to first attach to the target and then 
fold. The free-energy profiles associated with these two binding modes are 
compared.
\end{abstract}

\maketitle

%\newpage

\section{Introduction}

Protein structures are often viewed as well-defined static entities. 
For many proteins, this simplified static picture is accurate enough
to provide valuable information about the function.\cite{Branden:91} 
However, there are proteins that 
are wholly unstructured or only partially structured 
and yet functional. In fact, recent studies suggest that such proteins are 
more common than previously thought, especially in eukaryotic 
cells;\cite{Wright:99,Dyson:02} for instance, it has been estimated that as 
much as 17\% of the proteins in {\it Drosophila} are wholly 
unstructured.\cite{Dunker:01} In many cases, intrinsically unstructured 
proteins adopt specific structures upon binding to their biological 
targets.\cite{Wright:99,Dyson:02} Folding and binding are then coupled, 
thus establishing a direct link between folding and function.

It has been suggested that unstructured proteins offer several advantages 
in cellular regulation.\cite{Wright:99,Dyson:02} Being unstructured might,
for instance, allow one protein to interact with several targets. 
It might also be useful for control purposes due to 
rapid turnover. Furthermore, it has been argued that being unstructured 
might facilitate the binding of the protein to a target,\cite{Shoemaker:00} 
by increasing the `capture radius'. This mechanism, termed 
`fly casting',\cite{Shoemaker:00} was analyzed using a  
low-dimensional representation of the binding process. 
More recently, the thermodynamics of coupled folding-binding 
were examined using a generalized random energy model.\cite{Papoian:03} 

Here we study coupled folding-binding by computer simulations of a simple 
chain-based model. Simulating coupled folding-binding of a chain in a 
controlled manner requires proper sampling of the full conformational 
space and is harder than simulating folding of an isolated chain, 
since rigid-body translations and rotations must be taken into account. 
The complexity of the problem makes it highly desirable to study  
coarse-grained models before entering high-resolution modeling. 
For the present study, we use the minimal two-dimensional 
hydrophobic/polar HP lattice model of Lau and Dill.\cite{Lau:89} 
This model has been widely used to investigate basics of 
protein folding.\cite{Chan:02} This
and similar lattice-based models have also been used to gain insights 
into topics such as 
protein evolution,\cite{Bornberg-Bauer:97,Cui:02,Blackburne:03} 
prion-like conformational propagation\cite{Harrison:99,Harrison:01} and 
protein aggregation.\cite{Broglia:98,Istrail:99} 

Our study is inspired by recent experiments by  
Wahlberg~\etal\cite{Wahlberg:03} on the {\it in vitro} evolved protein 
\Zspa. This sequence was engineered\cite{Eklund:02} from the Z domain of 
staphylococcal protein A, a well characterized three-helix-bundle 
protein.\cite{Tashiro:97} It was selected for binding to the Z 
domain itself. In the Z:\Zspa\ complex,   
\Zspa\ adopts a structure similar to the solution
structure of the Z domain.\cite{Wahlberg:03,Hogbom:03} However,
in solution, \Zspa\ turns out to be structurally 
disordered.\cite{Wahlberg:03} The engineered \Zspa\ thus exhibits coupled 
folding-binding. In Ref.~\onlinecite{Favrin:03},
computer simulations of the solution behaviors of 
\Zspa\ and the Z domain were performed, using a relatively  
detailed off-lattice model with 5--6 atoms per 
amino acid.\cite{Favrin:03} Simulating the binding behavior of an 
unstructured protein like \Zspa\ 
at this level of resolution remains, however, a challenge.     

The present study consists of two parts. First, 
we study binding statistics for very short HP chains 
with up to $N=14$ monomers, to get an idea of 
how likely binding is to occur for stable and unstable sequences,
respectively, in this model.   
We then perform a more detailed study of two $N=25$ sequences.
In particular, this study allows us to compare the free energy of coupled 
folding-binding with free-energy profiles for docking and for
folding of an isolated chain.

\section{Model and Methods}
\label{sec:mm}

In the HP model,\cite{Lau:89} the protein chain is
represented by a string of hydrophobic (H) or polar (P) beads on 
the square lattice. Adjacent beads along the chain are connected by links 
of unit length. It is forbidden 
for two beads to occupy the same lattice site. 
Two beads that are neighbors on the lattice but non-adjacent
along the chain are said to be in contact. The energy that a 
configuration gets is determined by the number of HH contacts, 
each HH contact being assigned an energy $\epsilon<0$. This defines 
the model for a single chain in isolation. 

Here we study single chains interacting with some fixed target structure. 
Two types of targets are considered. In Sec.~\ref{sec:stat}, the 
target is an immobilized HP chain, and the inter-chain interactions 
are taken to be the same as the intra-chain interactions; that is, each 
HH contact between the target and the flexible chain is assigned the    
energy $\epsilon$. Any cross HH contact 
is given this energy  whether or not it is
present in the final bound state. In Sec.~\ref{sec:ex}, 
the target is an extended structure with one particular binding site. 
In this case, there is only one specific bead of the flexible chain that can 
interact with the binding site of the target. A contact between these 
two sites is taken to be favorable by an energy $\eb=3\epsilon$.           
 
To study the thermodynamics of these systems, we use Monte Carlo
methods. The moving chain is allowed four types of moves:
local one- and two-bead moves, non-local pivot moves, and one-step
translation moves (left, right, up or down). All moves are subject
to a Metropolis accept/reject step, to ensure that detailed balance 
is fulfilled. One round of simulation, called one `sweep', consists of 
(at most) $N-1$ one-bead moves, $N-2$ two-bead moves, one pivot move and 
one translation move, $N$ being the number of beads. 

In Sec.~\ref{sec:ex}, in addition to the thermodynamic simulations,
we also study the binding process as a function of Monte Carlo time. 
These calculations follow exactly the same protocol, except 
that the non-local pivot update is omitted, to 
avoid large unphysical deformations of the chain.  

With its simplified conformational space and its minimal two-letter 
alphabet, the HP model is not meant for studies of specific proteins, 
but rather to shed light on general questions for generic 
sequences. It is worth noting\cite{Chan:02} that if 
the principle of minimal frustration\cite{Bryngelson:87} holds, 
then the native structures of functional proteins should be strongly 
favored by the hydrophobicity pattern alone. This suggests that 
the HP model, despite its simplicity, might be able to capture non-trivial
features of the mapping from sequence to native structure. A statistical 
analysis of HP model sequences with unique ground states lends
support to this view; it turns out that the hydrophobic 
beads are anticorrelated along the chains,\cite{Irback:00,Irback:02} 
which is the same behavior that real (globular) protein sequences 
show.\cite{Irback:96,Irback:00} A recent study of the  
distribution of hydrophobicity in protein sequences can be found in 
Ref.~\onlinecite{Sandelin:03}. 
For a study of designed hydrophobic/polar copolymers with  
positive hydrophobicity correlations, see 
Refs.~\onlinecite{Khokhlov:98,Govorun:01}. 

The HP model has the advantage over high-resolution models that 
exact results can be obtained for short chains, by exhaustive 
enumeration of all possible states.  On the square lattice, all possible 
sequences with unique ground states (in isolation) have been 
determined for $N\le25$,\cite{Irback:02} along with the corresponding 
structures. 
Below, we make use of these results. Sequences having a unique ground
state will be referred to as stable.    
  
\section{Results and Discussion}

\subsection{Statistics for Short Chains}
\label{sec:stat}

To get an idea of how likely different binding behaviors are to 
occur in this model, we start by studying a large number of short 
sequences with up to $N=14$ beads. 
Each sequence interacts with some immobilized
target chain with the same length. 
All target sequences considered are stable    
and held fixed in their unique ground-state conformations.    
The two-chain system is contained in a box of size $2N\times2N$. 
The target is fixed in the middle of the box so that 
the moving chain can attack it from any side. We say that the 
moving protein binds to the target if there is a unique minimum-energy
configuration for the two-chain system. Whether or not this criterion
is met can be determined with high confidence by Monte Carlo methods 
for these chain lengths. As an operational criterion for binding 
we require that the minimum-energy configuration is unique
and is visited at least 10 times during the course of the simulation. 
To ensure that these visits to the bound state are `independent', a  
visit is counted only when the system comes to this state after 
going to some state which had at least 20\% higher energy than the
bound state.

We first test all possible sequences with a given $N$ for binding to one 
particular target conformation with that $N$, using  $N=10$, 11 and 12.
As target, we somewhat arbitrarily pick the first entry in the list of 
all stable ground states obtained 
in Ref.~\onlinecite{Irback:02}.\cite{footnote} 
The results of these calculations 
are summarized in Table~\ref{tab:1}. 
From this table it can be seen that the 
fraction of binding sequences is a few per cent, 
which is comparable to the number of stable sequences. A clear
majority of the binding sequences are unstable, 
which means that 
docking cannot be the mode of binding.
That most binding sequences are unstable does not mean that such 
sequences have an intrinsically higher propensity to bind than stable ones, 
but rather it merely reflects the fact that most sequences are unstable.      

\begin{table}[b]
\caption{\label{tab:1}
Numbers of sequences for $N=10$, 11 and 12 that can bind 
to the ground-state conformations
of HHPPHPPHPH, HHPPHPPPPHP and HHPPHPPHPHPH, respectively. 
The stable sequences for these $N$ are known from 
previous work.\cite{Chan:91,Irback:02}} 
\begin{ruledtabular}
\begin{tabular}{lrrr}
                                      &  $N=10$  &  $N=11$  &  $N=12$  \\
\hline
Total no. of sequences, $2^N$         &  1024    &  2048    &  4096  \\
Stable sequences${}^{\rm a}$          &     6    &    62    &    87  \\
Binding stable sequences${}^{\rm b}$  &     0    &     6    &    18  \\
Binding unstable sequences            &    20    &    44    &   140  \\
Total no. of binding sequences        &    20    &    50    &   158  \\
\end{tabular}
\end{ruledtabular}
\noindent
\begin{tabular}{l}
${}^{\rm a}$ A sequence is stable if it has a unique ground state in 
isolation.\\
${}^{\rm b}$ A sequence is binding if the system of moving chain
plus target has a unique ground state. 
\end{tabular}
\end{table}

\begin{figure}
\includegraphics{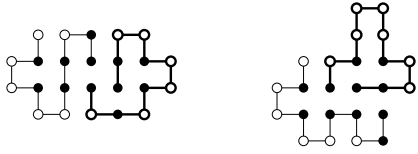}
\caption{\label{fig:1}
Same sequence (HHPHHPPHPPHP) binding to two different targets. 
The moving sequence (thin lines) adopts different folds as it binds to the 
two fixed targets (thick lines).}
\end{figure}

As mentioned in the introduction, one suggested use of being unstructured
is that it might enable the chain to bind to different targets. In the
model, there are sequences showing this behavior. An example of this
can be found in Fig.~\ref{fig:1}, which shows an unstructured $N=12$ sequence 
that is able to bind to two different targets (both of which are  
held fixed). The structure that the unstable sequence adopts upon binding is 
seen to depend on the target.   

Finally, restricting ourselves to stable sequences, we test for 
self-binding, by using the same sequence as target and moving chain 
simultaneously. In obtaining Table~\ref{tab:1} above, this 
calculation was done for those three sequences that served as targets. 
Using $N=12$ and $N=14$, we now consider all stable sequences. 
Table~\ref{tab:2} shows the results of these calculations. 
We see that $\sim$10\% of the stable sequences are self-binding both for 
$N=12$ and $N=14$. For $N=12$, we observe that all the 9 self-binding 
stable sequences bind in their ground-state conformations.  
For $N=14$, on the other hand, there exist 2 self-binding stable 
sequences for which the bound structure differs from the isolated one,  
while the remaining 40 keep the same structure after binding.

\begin{table}[b]
\caption{\label{tab:2}
Numbers of stable sequences that can bind to themselves 
for $N=12$ and $N=14$. The total numbers of stable 
sequences are from previous work.\cite{Chan:91,Irback:02}
The last row gives the numbers of self-binding stable
sequences that bind in their ground-state conformations.}
\begin{ruledtabular}
\begin{tabular}{lrr}
                                &  $N=12$ &  $N=14$   \\
\hline
Total no. of sequences          &  4096  & 16384   \\
Stable sequences                &    87  &   386   \\
Self-binding stable sequences   &    9   &    42   \\
Self-binding stable sequences with unchanged structure &  9 & 40 \\
\end{tabular}
\end{ruledtabular}
\end{table}

Let us briefly summarize the results presented so far.
Our study of all possible sequences for $N=10$, 11 and 12 shows that
a significant fraction (a few per cent) can bind to the targets 
considered. Furthermore, most of these binding sequences are 
unstructured in isolation.
Our study of all stable sequences for $N=12$ and $N=14$ shows that 
$\sim$10\% of them are self-binding. For most of these self-binding 
sequences, the complexed conformation is identical to the stable structure in
isolation, indicating the possibility of a docking-like binding behavior.

\subsection{Coupled Folding-Binding versus Docking}
\label{sec:ex}

\begin{figure}
\includegraphics{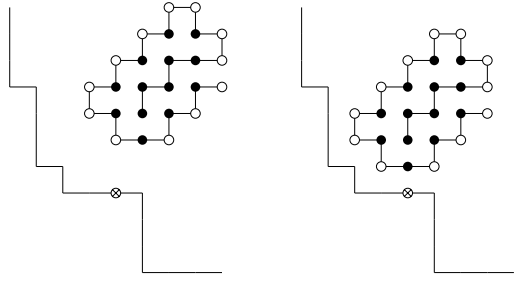}
\caption{\label{fig:2}
Typical binding behavior of sequence S. 
The chain first folds and then just translates to bind to the 
target. Filled and open circles represent hydrophobic and 
polar beads, respectively, and the circle with a cross represents the 
binding site.} 
\end{figure}

\begin{figure}
\includegraphics{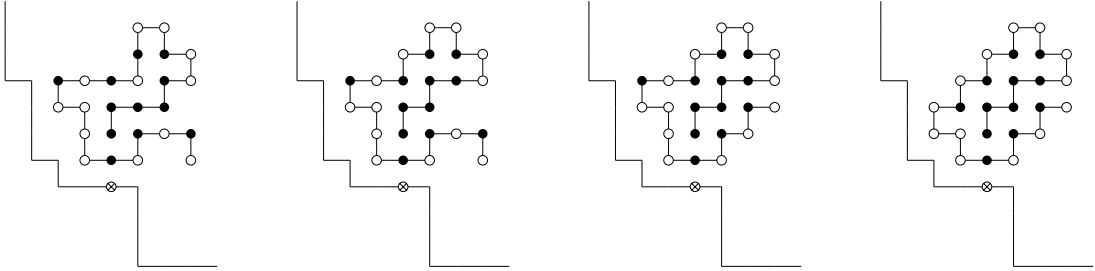}
\caption{\label{fig:3}
Snapshots from a simulation of sequence U. After binding to
the target, the chain rearranges itself into the minimum-energy 
state (right-most image). Symbols are as in Fig.~\protect\ref{fig:2}.}
\end{figure}

Having studied binding propensities of short chains, we now turn to a 
more detailed study of two longer chains with $N=25$, whose binding
behaviors are contrasted. This time the target is an extended structure
located in the left-bottom end of an $N\times N$ box, in which the 
chain is confined (see Figs.~\ref{fig:2} and \ref{fig:3}). 
The target has only one binding site, and the binding
is specific in that only one bead of the moving chain can interact 
with this binding site. 

The two sequences considered are given in Table~\ref{tab:3}. The
first sequence, called S, was studied in isolation in 
Ref.~\onlinecite{Irback:02}. 
It was obtained by applying a sequence optimization algorithm\cite{Irback:98} 
to 326  
stable sequences all having the same unique ground-state conformation. 
It turns out that S has a ground-state energy of 13$\epsilon$.  
The next most favorable conformations have 
only 11 HH contacts, so there is an energy gap
of $2\epsilon$, making S unusually stable. 
The other sequence studied, called U, is unstable. 
Its degenerate ground state has an energy of $11\epsilon$.
In our calculations, we observed five different 
conformations of U with this energy.
Sequence U is a close analog
of sequence S with just one hydrophobic bead mutated to a polar one.
The two sequences were taken to be similar in order for their bound 
structure to be the same.  

\begin{table}[b]
\caption{\label{tab:3}
The two $N=25$ sequences studied, S and U. The position at
which they differ is underlined. The bead that can interact with the target 
is in bold type.}   
\begin{ruledtabular}
\begin{tabular}{l}
S:\qquad HHHHH PPHPP HPHPH PP$\underline{{\rm H}}$P{\bf H} PHPHP \\
U:\qquad HHHHH PPHPP HPHPH PP$\underline{{\rm P}}$P{\bf H} PHPHP \\
\end{tabular}
\end{ruledtabular}
\end{table}

Our thermodynamic simulations of these two sequences were started 
from random configurations and contained $10^8$ Monte Carlo sweeps each.
The results show first of all that both these sequences do bind to the 
chosen target, in the sense that the system of target plus 
moving chain has a unique minimum-energy state. 
This state, referred to as the bound state,
was visited many independent times in the simulations. The bound
structure is the same for both chains (see Figs.~\ref{fig:2} and 
\ref{fig:3}), and coincides with the
stable structure of S in isolation. The energy of the bound state is 
$13\epsilon+\eb$ for S and $11\epsilon+\eb$ for U.

To characterize the binding behaviors of these sequences, we monitor the 
following two quantities:
\begin{enumerate} 
\item The binding parameter $I$. $I$ is the geometric distance between the 
binding site of the target and that bead of the moving chain that can interact 
with the target. As binding progresses, the value of $I$ reduces.
\item The folding parameter $\Qp$. The chains studied form 14 internal 
contacts, called native contacts, in their bound states. The number of these 
native contacts being present, $Q$, provides a measure of the `nativeness' 
of the chain. As the chain folds to its bound-state structure, the value
of $Q$ increases. It turns out that the value $Q=13$ is impossible to attain 
for these sequences. In our free-energy calculations, we therefore use a 
folding parameter $\Qp$ defined by  
\beq
\Qp=\left\{
        \begin{array}{ll}
             Q & {\rm if}\ Q\leq12\\
             13 & {\rm if}\ Q=14\\
        \end{array} \right.
\eeq
\end{enumerate}

\begin{figure}
\includegraphics[width=6.75cm]{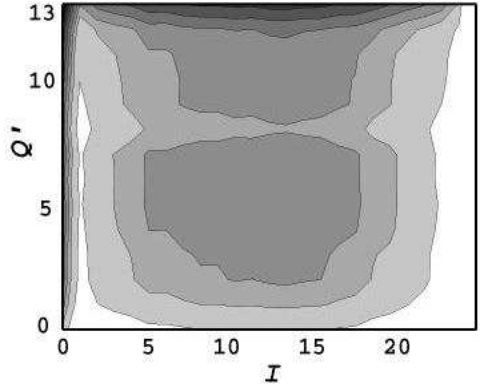}
\caption{\label{fig:4}
Free energy $F(I,\Qp)$ for sequence S at $T=\epsilon/2.6k$. 
The bound state, in which the chain is both folded and attached
to the target, corresponds to the upper left corner of the figure.
A chain that is folded but not attached to the target moves along the
upper edge of the figure (docking), whereas a chain that is attached to 
the target but incompletely folded moves along the 
left-hand edge (fly casting).  
The contours are spaced at intervals of 1\,$kT$
and dark tone corresponds to low free energy. Contours more than
8\,$kT$ above the minimum free energy are not shown.} 
\end{figure}

\begin{figure}
\includegraphics[width=6.75cm]{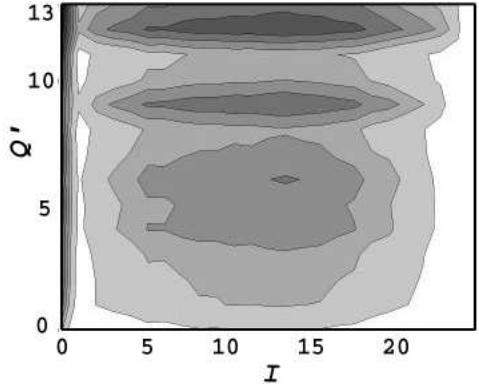}
\caption{\label{fig:5}
Same as Fig.~\ref{fig:4} for sequence U at $T=\epsilon/3.2k$.} 
\end{figure}

Figures~\ref{fig:4} and \ref{fig:5} show the free energy calculated 
as a function of these two variables, $F(I,\Qp)$, for the sequences 
S and U, respectively. The free energy is defined by 
$P(I,\Qp)=\exp[-F(I,\Qp)/kT]$, where $P(I,\Qp)$ is the joint 
$I,\Qp$ probability distribution. 
The temperature is taken as $T=\epsilon/2.6k$ for S and
$T=\epsilon/3.2k$ for U ($k$ is Boltzmann's constant). 
These temperatures are chosen such that the binding probabilities for 
the two chains are significant and close to each other 
($\approx 35$\,\%). Note that the systems are in their 
minimum-energy states if and only if $(I,\Qp)=(0,13)$. 

For sequence S (Fig.~\ref{fig:4}), 
we find that $F(I,\Qp)$ has
a simple shape with two narrow valleys along the lines $\Qp=13$ and 
$I=0$, respectively, and a broad and shallow minimum 
centered at $(I,\Qp)\approx(11,5)$, where the chain is unbound
and unfolded. This suggests that there are two very different major  
binding modes for this sequence. One way for the chain to reach 
its bound state is along the line $\Qp=13$, which corresponds 
to rigid docking; the chain first folds and then moves towards the 
binding site. The other major binding mode
is along the $I=0$ valley. Here the chain first attaches to the 
binding site and then folds to its final shape. Following 
Ref.~\onlinecite{Shoemaker:00}, we refer to this behavior as the fly-casting
mechanism. 

Let $\Pd$ and $\Pf$ denote the probabilities of finding the system
in the $\Qp=13$ and $I=0$ corridors, respectively
(not counting the bound state); that is,      
\beq
\Pd=\sum_{I>0} P(I,\Qp=13) \qquad\qquad \Pf=\sum_{\Qp=0}^{12} P(I=0,\Qp)
\label{corridors}\eeq
where the subscripts d and f refer to docking and 
fly casting.
For sequence S, we find that $\Pd$ is about twice as large as $\Pf$ (see
Table~\ref{tab:4}), suggesting that docking is the preferred 
binding mode for this sequence. 

\begin{table}[b]
\caption{\label{tab:4}
The probabilities $\Pd$ and $\Pf$ as defined by 
Eq.~(\ref{corridors})${}^{\rm a}$ 
along with the binding probability $\Pb=P(I=0,\Qp=13)$, for the sequences
S and U.}  
\begin{ruledtabular}
\begin{tabular}{lrr}
       &   S  &  U   \\
\hline
$\Pb$  & 0.35 & 0.35 \\
$\Pd$  & 0.32 & 0.05 \\
$\Pf$ & 0.17 & 0.23 \\
\end{tabular}
\end{ruledtabular}
\begin{tabular}{l}
${}^{\rm a}$ The subscripts d and f refer to docking and fly casting.
\end{tabular}
\end{table}

For the unstructured sequence U (Fig.~\ref{fig:5}), 
there is no 
free-energy valley corresponding to rigid docking, so $\Pd$ 
is small (see Table~\ref{tab:4}). There is, by contrast, 
an $I=0$ valley corresponding to fly casting for this sequence, too. 
The population $\Pf$ is, in fact, higher for U than for S 
(see Table~\ref{tab:4}). The $I=0$ valley is, for both sequences, 
separated from rest of the conformational space by a free-energy ridge. 
For sequence U, there are some narrow troughs in these hills towards 
the $I=0$ corridor, which should make it easier for this sequence 
to reach this corridor. In addition to the $I=0$  valley, sequence U exhibits 
two somewhat less pronounced valleys along 
the lines $\Qp=9$ and $\Qp=12$, respectively, as well as 
a broad unbound and unfolded minimum at lower $\Qp$. 
Note that there exist $\Qp=12$ conformations with minimal  
intra-chain energy for this sequence. The presence   
of the $\Qp=9$ and $\Qp=12$ valleys suggests that this chain can   
follow several different paths to its bound state. 
If the chain follows one of the fixed-$\Qp$ valleys, it reaches the 
binding site without having its full bound-state structure; 
folding is completed after the chain has attached itself to the target. 
When increasing the box size to $3N\times 3N$, 
we observed valleys at the same values of $\Qp$
which were stretched along $I$. 

\begin{figure}
\includegraphics[width=7cm]{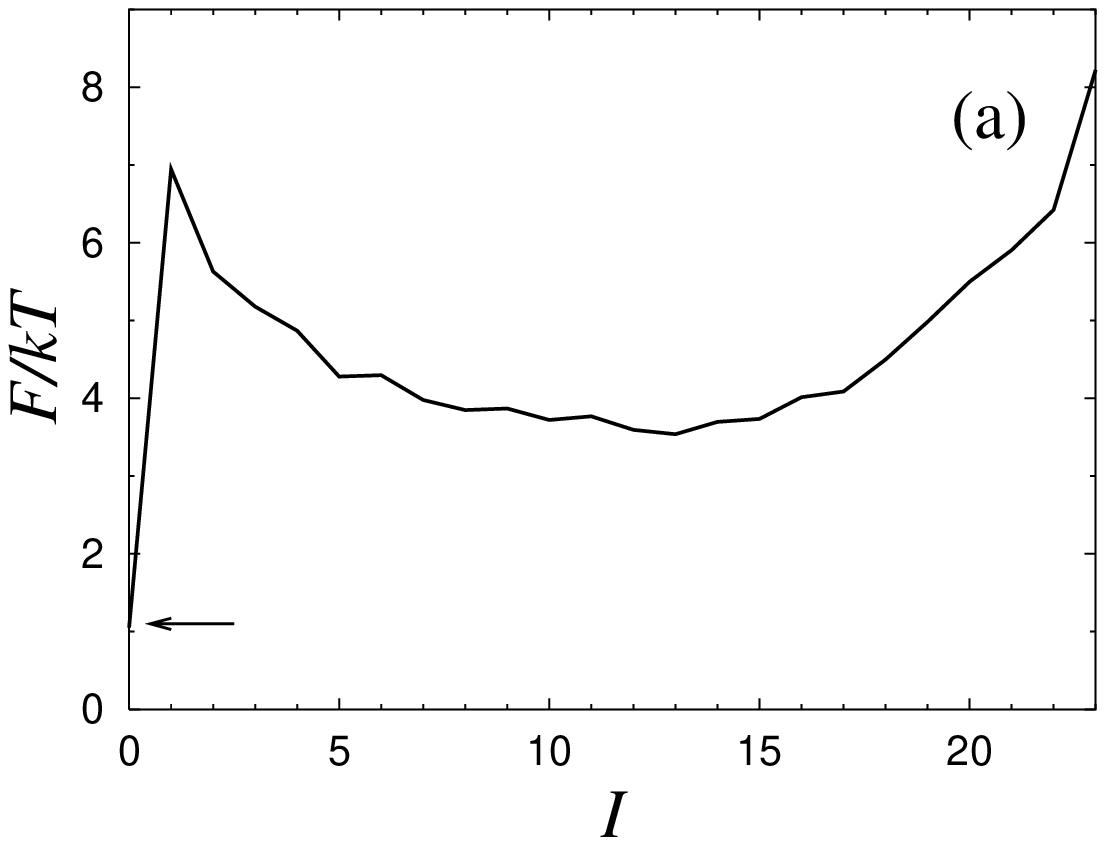}
\includegraphics[width=7cm]{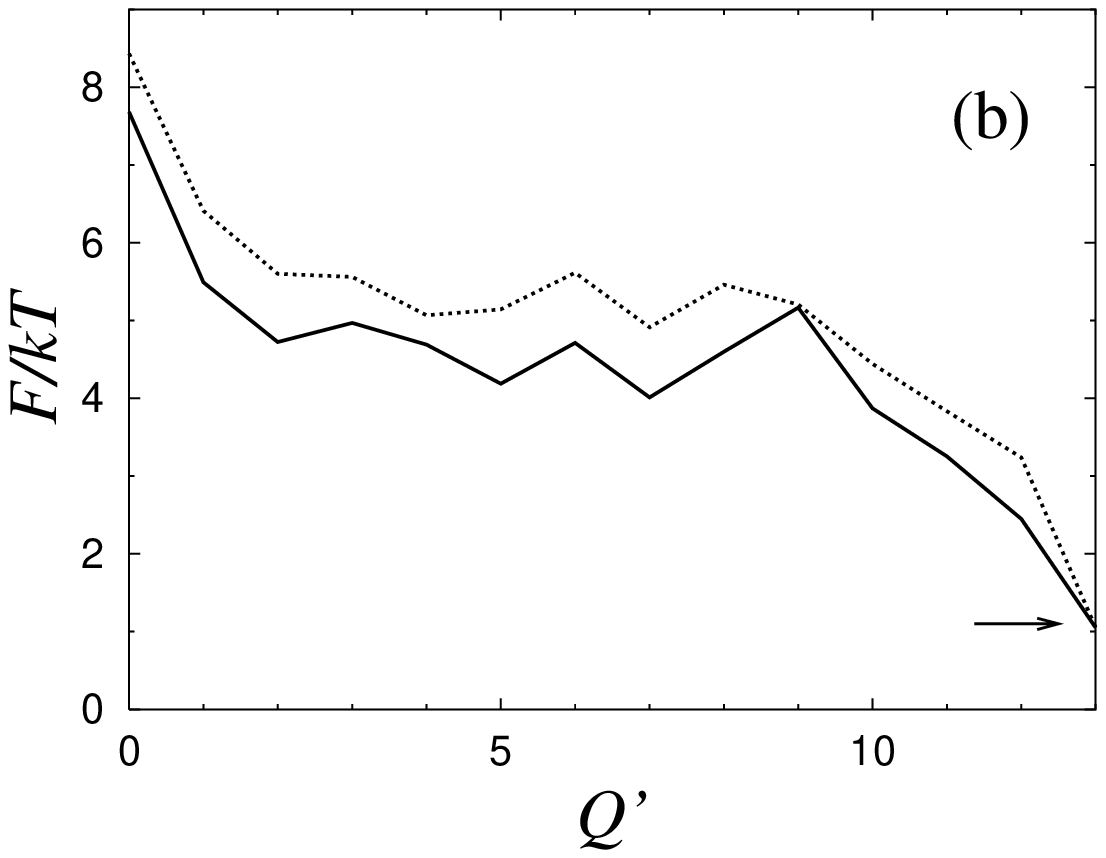}
\caption{\label{fig:6}
Free-energy profiles for the docking and fly-casting corridors. 
(a) $F(I,\Qp=13)$ for sequence S. (b) $F(I=0,\Qp)$ for sequence 
S (dotted line) and sequence U (solid line). 
The arrows indicate
the location of the bound state, in which the chains are both 
folded and attached to the target.
The temperature is $\epsilon/2.6k$
for S and $\epsilon/3.2k$ for U (same as  
in Figs.~\ref{fig:4} and \ref{fig:5}, respectively).}
\end{figure}

It is instructive to take a closer look at the free energy 
in the docking and fly-casting corridors, respectively. Figure~\ref{fig:6}a 
shows the free energy along the docking corridor, $F(I,\Qp=13)$,
for sequence S. We see that in order to reach the bound state, the 
chain has to pass a free-energy barrier with a height of $\approx$\,3\,$kT$.  
This barrier is entropic. Figure~\ref{fig:6}b shows 
the free energy along the fly-casting corridor, $F(I=0,\Qp)$, for
both sequences. Two features are worth noting. First, the shape of the
curve is roughly the same for the two sequences, although the curve 
is shifted downwards for sequence U which spends more time in this corridor. 
Second, there is no major free-energy barrier in this corridor
(the highest barrier is about 1\,$kT$), and the last part towards the
bound state (from $\Qp=8$--9 to $\Qp=13$) is downhill in free energy. 

\begin{figure}
\includegraphics[width=7cm]{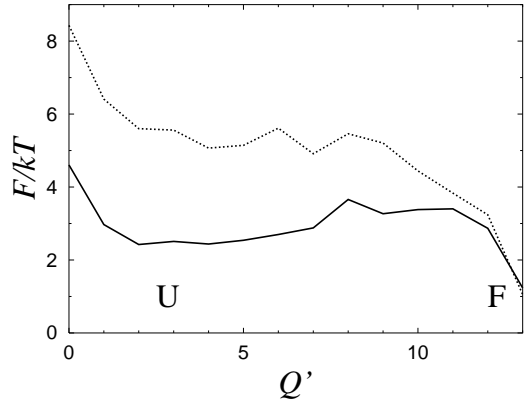}
\caption{\label{fig:7}
Two free-energy profiles for sequence S. The dashed line shows 
the free energy along the 
fly-casting corridor, $F(I=0,\Qp)$, at $T=\epsilon/2.6k$ 
(same as in Fig.~\protect\ref{fig:6}b). The full line shows $F(\Qp)$ for 
the isolated chain at $T=\epsilon/2.268k$. 
U and F refer to unfolded
and folded conformations, respectively.}
\end{figure}

It is also interesting to compare the free energy in the fly-casting 
corridor (Fig.~\ref{fig:6}b) with the free energy of folding for sequence S in 
isolation. For this purpose, we performed a simulation of S in isolation
at $T=\epsilon/2.268k$, where the native population (29\%) is comparable 
to the binding probability at $T=\epsilon/2.6k$ (35\%). From  
Fig.~\ref{fig:7}, it can be seen that the free energy for the isolated 
chain is markedly different from that for the fly-casting corridor.   
In particular, we see that the bound-state minimum of the interacting chain
is broader than the native minimum of the isolated chain; 
the isolated chain must reach $\Qp=11$ before the free 
energy starts to decrease. 

We also performed Monte Carlo-based kinetic simulations 
(see Sec.~\ref{sec:mm}) of the sequences S and U, 
using the same temperatures as in the 
thermodynamic runs. For each sequence, we carried out a set of 500 
simulations, each containing $5\times 10^6$ Monte Carlo sweeps. 
The simulations were started from configurations obtained by short 
runs at a higher temperature ($T=\epsilon/2k$). 
The resulting configurations from these preparatory runs (the
starting configurations for the kinetic runs) showed a wide variation 
in both location and conformation for the moving chain.  
The averages of $I$ and $Q$ over this ensemble were 
about $12$ and 4--5, respectively.  
 
\begin{figure}
\includegraphics[width=7cm]{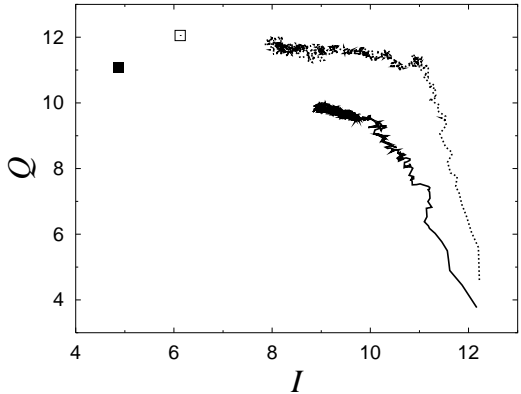}
\caption{\label{fig:8}
Parametric plot of $Q$ against $I$ with Monte Carlo time
as a parameter for sequence S (dotted line) and sequence U (solid line). The 
curves represent averages over 500 simulations and the
bottom-right ends correspond to time zero. 
The unfilled and filled squares represent the   
equilibrium points ($\ev{Q}$,$\ev{I}$) for S and U, respectively.
Temperatures are as in Figs.~\ref{fig:4} and \ref{fig:5}.}
\end{figure}

Figure~\ref{fig:8} shows the evolution of this ensemble of
500 systems with Monte Carlo time in the ($I,Q$) plane (not $\Qp$). 
This parametric plot, with Monte Carlo time as a parameter, has the 
advantage of being independent of the difference in time scale between the two 
different temperatures used. To reduce noise, each data point 
represents an average over $10^4$ Monte Carlo sweeps, 
thus giving 500 data points in total. 
In Fig.~\ref{fig:8}, we also indicate the equilibrium values of 
$I$ and $Q$, which were obtained by a separate, very long 
simulation. The equilibrium values are given by   
$(\ev{I},\ev{Q})=(6.1,12.1)$ for S and 
$(\ev{I},\ev{Q})=(4.9,11.1)$ for U. Note that, as expected,  
S has a higher value of $\ev{Q}$ and U has a lower value of $\ev{I}$.  

For sequence S, we see that the relaxation towards the equilibrium 
point has a clear two-step character, where the first step
corresponds to folding (increasing $Q$) and the second step  
to binding (decreasing $I$). This behavior is indeed what one 
expects if docking is the preferred binding mode. 
For sequence U, there is no such clear distinction between
folding and binding; $I$ and $Q$ evolve in a more correlated manner.  
Hence, the results of the kinetic simulations support the 
conclusions from the free-energy analysis.
    
\section{Summary}

We have studied different binding mechanisms of polypeptide chains
using the minimal HP model. The aim of this study is to understand
the nature of an important aspect of protein-protein recognition and
it can serve as a guideline for further, detailed studies of this complex
phenomenon. 

For small chain lengths, we found that the fraction of sequences 
that can bind to a given (fixed) target is comparable to the fraction of 
sequences that are stable in isolation. The overlap between these two sets
of sequences is small; most binding sequences are unstable in isolation 
and fold upon binding to the target.    
 
We then compared the binding behaviors of two related $N=25$ sequences. 
One of the sequences, S, had been optimized for high 
stability in isolation,\cite{Irback:02} and was found to have rigid
docking as its preferred binding mode. The other, slightly mutated 
sequence, U, was found to have a bound structure identical to
that of S, but to be unstructured in isolation. In these respects, 
sequence U is reminiscent of the engineered protein \Zspa\ mentioned
in the introduction. It turned out that sequence U tends to
first attach to the target and then fold, which supports the
conclusion\cite{Shoemaker:00} that unstructured chains prefer binding 
through a fly-casting mechanism. The free-energy profile associated with
this binding mode was found to lack high barriers and to exhibit a 
broad bound-state minimum, compared with the native minimum of sequence 
S in isolation. The observed difference in shape between these free energies 
suggests that by attaching itself to the target, 
the chain becomes able to fold more efficiently than it does in 
isolation. The free energy of docking obtained for sequence S contains, 
as expected, a significant entropic barrier, which the chain has to overcome 
in order to reach the bound state.      

These free-energy profiles were obtained using two particular sequences 
in this model. Therefore, it should be stressed that the shape of the 
free energy was not considered at all when selecting these two sequences
and the target structure; the goal was just to have two binding sequences 
with the same bound structure but different stability properties in isolation.
As a result, we believe that the trends seen have some generality. 
Nevertheless, it is clear that it would be highly desirable to extend
these calculations to other and longer sequences and to more
realistic off-lattice models. Furthermore, it would be interesting 
to study the kinetics over a larger time interval; this requires a  
larger number of systems than 500 which we used, in order to keep
the statistical errors under control.     

\subsection*{Acknowledgments}

We thank Bj\"orn Samuelsson for valuable discussions and help 
with figures. 
This work was in part supported by the Swedish Research Council.

\end{document}